# Won't you see my neighbor?: User predictions, mental models, and similarity-based explanations of AI classifiers

Kimberly Glasgow, Jonathan Kopecky, John Gersh, Adam Crego,
Johns Hopkins University Applied Physics Laboratory

Humans should be able work more effectively with artificial intelligence-based systems when they can predict likely failures and form useful mental models of how the systems work. We conducted a study of human's mental models of artificial intelligence systems using a high-performing image classifier, focusing on participants' ability to predict the classification result for a particular image. Participants viewed individual labeled images in one of two classes and then tried to predict whether the classifier would label them correctly. In this experiment we explored the effect of giving participants additional information about an image's nearest neighbors in a space representing the otherwise uninterpretable features extracted by the lower layers of the classifier's neural network. We found that providing this information did increase participants' prediction performance, and that the performance improvement could be related to the neighbor images' similarity to the target image. We also found indications that the presentation of this information may influence people's own classification of the target image— that is, rather than just anthropomorphizing the system, in some cases the humans become "mechanomorphized" in their judgements.

## INTRODUCTION

Effective human-machine teaming depends on interpredictability (Johnson, et al. 2014, Klein, at al. 2005). From the human viewpoint, this requires that people be able to make reasonable predictions of machine behavior and performance. Understanding limits to the machine's performance is particularly important for teaming (Bansal, et al., 2019). Such understanding is crucial both to teaming performance per se and to the proper calibration of human trust in the machine partner (Tomsett, et al., 2020). Machine learning- [ML]-based systems are problematic in this regard; it is too often difficult for people to predict, for example, a classifier's response to a particular stimulus or class of stimuli. Making such systems interpretable of explainable is an active area of research. At the same time, though, it is important to understand how people develop mental models of an ML-based system's operation. Such understanding can inform the design of such systems, both for operation and for interactive training (Amershi, et al. 2014). We are conducting research into the development of mental models in this context. Earlier work investigated people's ability to predict an image classifier's performance and their development of a mental model of its function (Bos et al., 2019). We found that, in our experimental context, people did not learn to predict classifier performance over several trials. We also found that they tended to attribute to the machine classifier aspects of their understanding of their own visual and perceptual processes.

In the research reported here, we provided participants with additional information about classifier behavior through the presentation of additional images whose features (according to the classifier) are similar to an image for which they are asked to make a classification prediction. We explore whether this additional context improves participant accuracy or changes their mental models of the classifier.

This research uses a machine-learning-based image classifier, which is a relatively familiar and approachable artificial intelligence system. ML-based image classification has shown tremendous progress, particularly after developments in convolutional neural networks such as Google's Inception (Szegedy et al., 2014). Because such systems are very complex and have no human-like semantic preconceptions or other knowledge of the world, it can be very difficult to know how a trained neural network comes to the conclusions that it does.

Complete understanding of such systems by an individual difficult if not impossible. Even their designers, who understand how the original networks were trained, can only infer how a self-trained system that has been exposed to thousands of images will respond to a novel image. Research on how to help humans make sense of these systems via various means is an active area of research (see Olah, et al. 2018 for a brief review.).

We consider a different approach here. We ask whether untrained users can construct useful mental models to understand and predict these systems' performance. Humans have impressive powers of inference and their own highly developed perceptual systems, and with enough experience might be able to predict performance without fully understanding the underlying mechanism. There are open questions about whether this is possible, how much experience would be necessary, and what kinds of analysis and feedback would optimally support this learning. This study considers one type of additional context to support learning, exposure to images similar to one on which classifier performance is to be predicted. In particular, we present participants with images that are close neighbors of the subject image *in a space which represents features known to be important to the classifier.* Other research has explored presenting users with a spatial layout of similar images, but in the context of trust calibration rather than mental model investigation and depicting custom layouts rather than feature-space 1-D or 2-D maps (Yang, et al. 2020).

Another set of questions relates to the contents of human mental models. What do users understand, or think they understand, about image classifiers? The answer to this is unknown. There has been little prior research on human mental models of image classifiers. Prior research has examined mental models of other complex engineered systems, some of which have involved machine learning (Tulio, et al. 2007, Kulesza, et al. 2012). This research has in general sought to understand the content and structure of mental models, but has not linked that to human-machine task performance. A mental model is a cognitive representation of some aspect of the world. In a functional sense, all mental-model theories involve the prediction of some aspects of human behavior through a conceptual framework involving something called a mental model. People do form mental models of systems, even opaque systems, or at least respond to questions indicating so. (Johnson-Laird, 2004; Norman, 1983). Mental models are used to make predictions. (Rouse & Morris, 1986). Developing a mental model involves both induction and *abduction* – perception and *explanation.* (Khemlani & Johnson-Laird, 2011; Klein, Phillips, Rall, & Peluso, 2007)

**Research questions:**
1. Can untrained humans improve their predictions of image classifier success after one session of practice and informative feedback?
2. Can additional context, in the form of similar images (neighbors in classifier feature space) further improve predictions of image classifier success?
3. What mental models of image classifiers do humans construct to help with this predictive task? Will providing additional context lead to different mental models?

## METHODS

We pursued these questions with an online experiment in which images are classified as related to the game of baseball or not. Participants were asked to predict whether the image classification algorithm would correctly label given images as baseball-related or not. The first two research questions were addressed by a quantitative analysis of participants' prediction success, the third by a qualitative analysis of comments made by participants during the prediction task.

### Image Classification Dataset and Model

The dataset included 9023 images, of which 4,750 images related to baseball and 4,273 images were of images that were not but typically involved other sports. The baseball images involved many facets of the game. These included images of professionals or amateurs playing baseball, baseball stadiums, baseball fields, baseball cards, and so on. Most of the not-baseball images similarly portrayed other sports, athletes, and venues, including soccer, basketball, lacrosse, and others.

The classifier combined a pre-trained convolutional neural network (CNN) with a random forest classifier (Rodriguez, et al. 2014). The CNN was trained on the Imagenet collection of images of all kinds. The final ten layers were removed, leaving a network which performs featurization somewhat similar to human bottom-up perception. Using the pretrained network greatly decreases training time and reduces the number of images needed. The shortened CNN does not change with training on images to be classified, but performs featurization of those novel images. These features (a 4,096-length vector) are then fed into a random forest classifier and trained with images in the set of interest (in this case of baseball and other sports). The random forest classifier can be trained relatively quickly and attains over 90% accuracy. However, as images for use in the experimental task were specifically selected based on their class (baseball or not baseball), the distribution of neighbor images, and whether their classification was correct, this metric does not capture the properties of the experimental stimuli.

In order to determine image similarity, we performed dimensionality reduction on the 4096-component feature vector using t-distributed stochastic neighbor embedding , or t-SNE (Krijthe van der Maaten, & Krijthe, 2018). This reduced the image representation to 2 dimensions. This allowed us to generate a "map" of all the images, based on their t-SNE coordinates, and to determine the nearest neighbors of an image based on the Euclidian distance between them. See Figure 1. Many images had nearest neighbors that could be of a different category than the target image.

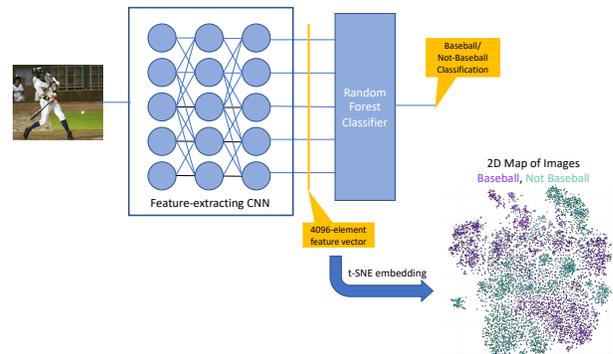

*Figure 1: Illustration of image classification and dimensionality reduction approach*

### Experimental Task

Participants viewed and made predictions of classifier results for a total of 60 images across three blocks of 20. In the control case, participants were shown only the classification target image. In the experimental case, they were shown the classification target and its four closest neighbors in the 2-D t-SNE space. Participants were also asked for their own classification of the image, asked to comment on the reason for each prediction, and to rate their confidence in their predictions. They were also asked general questions after each trial block. The study was approved by the Johns Hopkins School of Medicine IRB (00174285).

### Participants

Participants were provided by Amazon Turk. There were 149 participants (75 female, 74 male); they received $15 compensation, plus a $5 bonus to the top 35% of performers on the final block. The mean age was 42, and ranged from 24 to 69. Most participants had completed college (55%), with an additional 30% having some college and 13% with a high school diploma. Some participants (11%) had had a major in college involving computers (either computer science of computer information systems). Most participants had no experience with image classifiers (66%), with a few (3%) having had some experience outside of Turk and 6% having had previous experiences with them in Turk.

**Procedure**

The experiment was administered in Qualtrics, with all participants being given 3 blocks of 20 trials each. On each trial, participants would see an image and then be asked first if they believe the image was about baseball in some way, and then asked if the baseball classifier would classify this image as about baseball in some way or not, rate their confidence in their prediction, and to describe their reasoning behind their prediction (This was free text, and was optional on any given trial but had to be done at least 5 times per block.). After each trial, the participant received feedback in the form of the actual classification for the image. Participants in the control condition saw single images throughout the 3 blocks, while participants in the experimental condition saw the image as well as its four closest neighbors in a row beneath it in the second two blocks (Fig. 2) Note that target images were the same in both conditions, and that the order of the second and third blocks were counterbalanced.

The stimuli for each block were chosen so that there were 10 images of baseball and 10 images not of baseball, and the classifier was correct on 14 of the 20 images. The six incorrect classifications were equally divided between false positives and false negatives. Chance performance was thus 70% accuracy, if the participant predicted that the classifier would always answer accurately. The stimuli were also designed so that the nearest neighbors would often be of a class different than the target image.

**Comment analysis**

In order to investigate participant's mental model development, we coded their text responses, using the Taguette qualitative analysis tool (Rampin, et al., 2021). We developed a codebook based on prior experiments (Bos, et al. 2019) and early pilot studies. It included 46 different tags ranging from mentions of individual items in the image (e.g., sports equipment), to the presence of people (e.g., players, fans), to action or motions, to the venue (e.g., playing field, stadium), to more abstract (e.g., an overall baseball-like gestalt). A subset of participants' responses was tagged by at least two coders.

## RESULTS

**Prediction of machine classification**

The major quantitative objective of the experiment was the accuracy of participants' prediction of the machine's classification. There was no difference in accuracy in the first block between conditions (mean control = 12.0, mean experimental = 12.3, t(147) = 0.37, p=ns). This was expected, since the task was the same for the two conditions. In the second block, participants in both conditions improved but participants in the experimental condition did significantly better (mean control = 13.4, mean experimental = 13.9, t(147) = 1.99, p < .05). By the third and final block, the difference was even more pronounced, with the control group having a mean of 12.8 and the experimental group having a mean of 14.3 correct (t(147) = 5.53, p < .0001). There was a significant effect of both condition across the 3 blocks (p<.001) as well as a significant interaction between condition and accuracy F(2,2) = 6.5, p<.01.

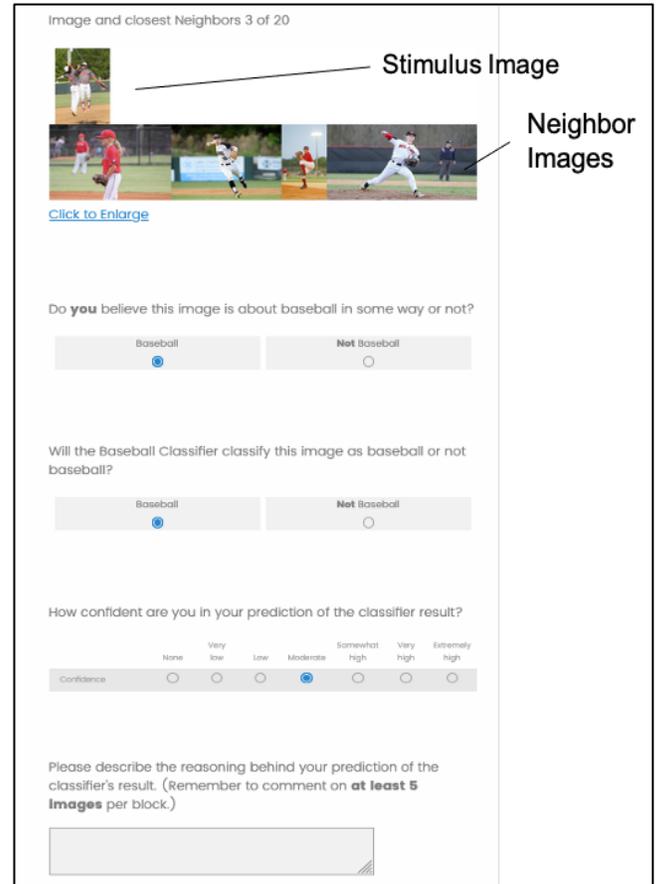

*Figure 2: Example of an experimental stimulus with target image on top and its four nearest neighbors below it. Participants stated their own classification of the image, predicted what the classifier would do, provided their confidence of that prediction, and had an opportunity to comment.*

Did this improvement relate in any way to participant's use of the depiction of the target image's nearest neighbors? Fig. 3 depicts a scatter plot comparing prediction performance improvement to the content of the nearest-neighbor images. Each point represents an image presented to the participants in the second two blocks. The horizontal axis represents the improvement in mean prediction performance (experimental – control) for the image. The vertical axis represents the fraction of the four neighbors of that image which have the same classification as the classifier result. For example, if the classifier result is baseball, and three of the neighbors are baseball images, then the fraction is 0.75.

The plot indicates that participants are using the information in the neighbor images in making their predictions. In general, participants in the experimental condition do better when more of the neighbors match what the classifier will say. When fewer of them do, they do worse.

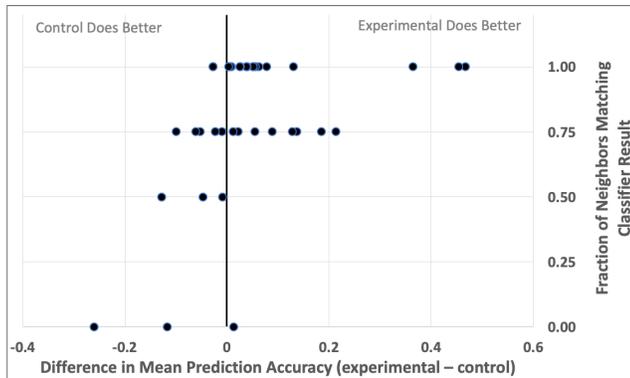

Figure 3: Scatterplot of prediction performance improvement in the experimental condition vs. the fraction of neighbor images matching the classifier result for the target image. Each point represents an image in the second two blocks.

**Confidence**

Participants rated their confidence in their prediction of the classifier's result for each trial from 1 (None) to 7 (Extremely High Confidence). We removed the first block from all subsequent analyses, because in that block participants are first calibrating their confidence as they learn more about the classifier. It turned out that more information lowered confidence overall, as control participants mean confidence (4.89) was significantly higher than the mean experimental group (4.75, t(5958) = 3.84, p<.001). There was a significant relationship between reported confidence and accuracy over the final two blocks (from a logistic regression, the regression coefficient was 0.12, p < .001) along with a significant main effect of condition (coefficient = -0.12, p<.001), indicating that at a given level of confidence, control participants were significantly less accurate than experimental participants. These results hold if we examine only the final block for each participant as well. However, the interaction between condition and accuracy was not significant (Confidence coefficient of 0.13, and condition effect of -0.18, both p<.001, but interaction of -0.01, p=NS). See the figure below for accuracy over the final 2 blocks. Further research may be needed to investigate whether these data represent the well-known overconfidence effect (Kahneman et al., 1982).

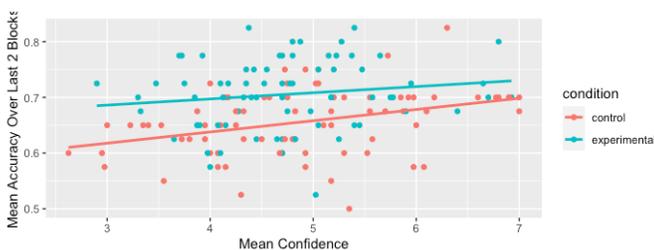

Figure 4: Participant accuracy and confidence over the final two blocks

**Participants' image classification errors**

Participants were not always correct in their own classification of the target images. Unexpectedly, we found a significant difference in ability to determine the true class of an image between the control group and the experimental group. Participants in the control group got about 98% of images correct in all three blocks, and had an overall accuracy of 98.1%. Participants in the experimental condition, however, performed significantly worse. Their overall accuracy was 94.9%, but this is an underreporting of the magnitude of the difference because in the first block (where they saw the same stimuli as participants in the control condition), their accuracy was 98.9%. By their second block, experimental participants only got 92.4% correct and in their third block, they got 93.4% correct. All of these differences were highly significant (p<.001). This suggests that providing the context of an image in classifier feature space may influence how that image is perceived. In a sense, this may be the inverse of the effective anthromorphism reported earlier (Bos, et al. 2019, Mueller 2020); people may "mechanomorphize" themselves.

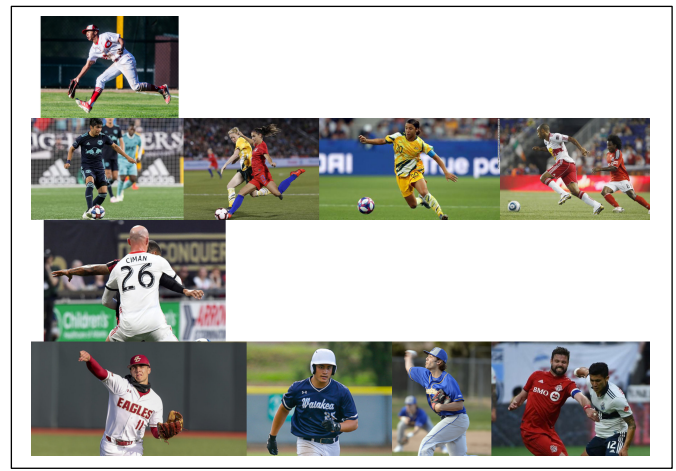

Figure 5: Two stimuli for which the experimental group were less accurate at determining the correct class of the stimulus image (73% and 56%, respectively, vs. 99% accuracy for control).

**Participant comments**

Participants exhibited some differences in their willingness to comment on a particular trial. Participants had to comment on a minimum of 15 trials across blocks (at least 5 per block), and could comment on as many as 60 trials across blocks (20 per block). Across conditions, 11 participants commented the minimum number of times (5 in control condition, 6 in the experimental condition). Thirty participants commented on every trial across blocks (19 in the control condition, 11 in the experimental). The overall mean of comments across participants was 59% of trials with a median of 50%. If we exclude the first block because participants were more likely to comment there due to the novelty of the task and being extra compliant, these numbers went down slightly, with a mean of 57% and a median of 47.5% of trials having comments.

Participants were more likely to comment if they were in the control condition. With the first block included participants in the experimental condition commented on 55% of trials, those in the control condition commented on 62%, p <.001) With the first block excluded (control participants commented on 61% of trials, experimental participants commented on 54%, p<.001).

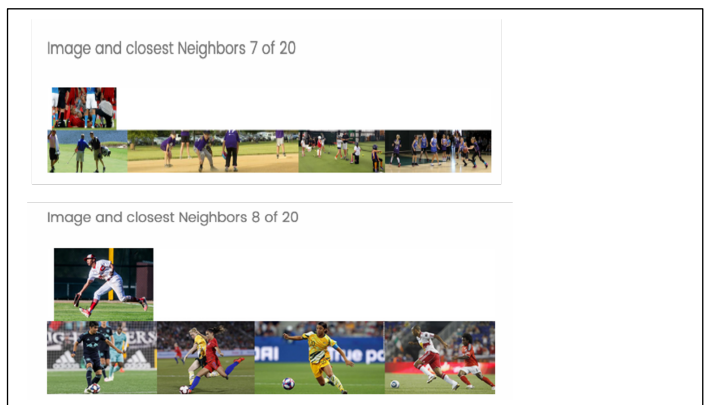

*Figure 6: Stimulus 7, received more comments from control vs. experimental subjects. The opposite pattern was observed for stimulus 8.*

Participants were most likely to comment on trials in which the classifier incorrectly asserted that a non-baseball image was about baseball, both across blocks and even excluding the first block ($\chi^2(3) = 120, p < .0001$). These false positives were the most commented about type of trials. In the design of the experimental trials, participants comment *prior* to learning the decision of the classifier; thus, the elevated rate of commenting is not in response to classifier error.

The number of times participants commented on false positives was also significantly greater than the number of times participants commented on false negatives, where the classifier incorrectly classified a baseball image as being not about baseball. Participants were also more likely to comment when they predicted that the classifier would make a different classification than they did, (*discordant* predictions) (61% of the time compared to 58% of the time when their classification was the same). This effect was larger if we exclude the first block, as participants developed their understanding of the classifier (64% of the time they commented on discordant trials compared to 56% of the time when they agreed.) These differences were statistically significant.

Consistent with the overall commenting statistics, on most trials, participants in the control condition were more likely to comment on that trial than were the experimental participants. Trial 8 of Block C was an exception, with experimental participants commenting 33% of the time compared to 28% for control participants. But the previous trial, Trial 7 of Block C, had the biggest difference overall, with control participants commenting 33% of the time, but experimental participants commenting only 23% of the time. Participants' willingness to comment on false positive results may indicate that these results triggered reflection on classifier processing and possible adjustments of their mental models. Further research would be needed to investigate this further.

*Table 1: Rates of participant commenting by classifier result (type of stimulus)*

| Classifier Result | % Time Commented (all blocks) | % Time Commented (no block A) |
|---|---|---|
| True Positive | 59% | 52% |
| True Negative | 53% | 54% |
| False Positive | 70% | 72% |
| False Negative | 61% | 63% |

**Qualitative aspects of user mental models**

Certain themes emerged from the qualitative analysis. Participants often commented on the images in terms of:
- Sport or sport-related (e.g., the image as being about baseball or not being about baseball)
- Players or people, such as discussing a baseball player or the crowds
- Equipment (e.g., discussing a bat or glove or the ball itself)
- Venue (field or stadium)
- Image properties (e.g., field of view or blurriness in the image)
- Actions, positions, or poses (e.g., describing a person as being in a pitching position)
- How well the image fit existing human domain knowledge or their mental models (e.g., was the image consistent with the idea of baseball or was there a mismatch?)
- Evaluative comments, such as:
    - Presence of absence of key features,
    - Prior behavior of the classifier
    - Overall gestalt sense of the image
    - In the experimental condition, comments about the neighbors' similarity to the stimulus image or to each other

The comments ranged from short, uninformative comments (e.g., "not sure") to comments expressing sophisticated mental models (e.g., "The [soccer] ball is much larger than any ball you would typically see in a baseball image."). Overall, there were about 8,000 comments comprising nearly 100K words. Tagging all these comments would not have been feasible, so we prioritized participants who did well in the final block, especially when they predicted the classifier would make a different prediction than they did (discordant predictions), as well as participants who performed poorly when they made discordant predictions. Lower performers struggled to identify any pattern in why the classifier got some right or wrong (e.g., "Understanding? The only thing that changed over the course of the experiment was my confusion."). Higher performers seemed to adapt a bit better as they tried to discern what the classifier was doing (e.g., "My ideas about what the classifier would be good at identifying changed. Instead of bats and gloves, it turned out the field and players uniform seemed to be more critical.")

Across conditions, participants in the experimental condition commented more on the classifier itself, making more comments about it in Blocks 2 and 3 than did control participants, despite no difference in the first block. Perhaps seeing related images makes the classifier classification itself more salient, and participants discuss what it is doing *across* images rather than discussing the stimulus image *itself* in quite as much detail. This could also explain why comments on the 'utility' (explaining why something is useful to the classifier) also differed between experimental and control participants. Conversely, control participants were more likely to comment on equipment and uniforms than were experimental participants. They also were more likely to discuss the gestalt of the image (e.g., saying an image is 'clearly baseball').

In the experimental condition, the class of the neighbors also affected the likelihood of commenting. This likelihood was not symmetric. When all the neighbors were baseball-related, participants commented 62% of the time. But when all the neighbors were not about baseball, participants commented only 51% of the time (this was statistically significant, p<.001). The table below shows the percentage of comments amongst the experimental participants only for blocks where they saw neighbors.

*Table 2: Neighbor classes and likelihood of commenting*

| Neighbor Class | Commented (%) |
|---|---|
| B, B, B, B | 62% |
| B, B, B, NB | 60% |
| B, B, NB, B | 36% |
| B, NB, B, B | 41% |
| NB, B, B, B | 68% |
| NB, B, NB, NB | 27% |
| NB, NB, B, NB | 64% |
| NB, NB, NB, NB | 51% |

## DISCUSSION AND CONCLUSION

Humans can work more effectively with powerful AI systems when they can predict likely failures and form useful mental models of how those systems work. To study this, we asked human participants to try to predict the outcomes from a high-performing but quite opaque image classifier. This was a short task (60 images total) and participants had to form whatever mental models they would use quite quickly, without being able to test a large number of hypotheses about what cues the classifier might be using. Participants in the experimental condition, who were exposed to additional context (neighbor images) significantly improved the accuracy of their predictions. We found in addition that the magnitude of this improvement depended on the similarity of the neighbor images to the target image.

Presenting more information by presenting similar images to the image to be classified had both expected and unexpected effects. One expected effect is that is provides more information, so participants are more likely to accurately assess what the classifier will do, and we find evidence of this. One unexpected finding is that the image could influence how the human perceives the image, making them actually worse at classifying the image than without that extra information. Another unexpected finding is the way that this extra information changes how a participant will comment on their prediction of the classifier's result. Participants with less information tend to discuss image details, whereas those with more seemingly tried to find a pattern across the images and so focused more on the classifier itself and less on the image.